\documentstyle[11pt]{article}
\newfont{\tenmsb}{msbm10 scaled\magstep1}

\let\ssection=\section\renewcommand{\section}{\setcounter{equation}{0}\ssection}

\def\smallover#1/#2{\hbox{$\textstyle{#1\over#2}$}}

\input{tcilatex}
\begin{document}

\title{Deformation of surfaces, integrable systems and Self-Dual
Yang-Mills equation}
\author{T.A.Kozhamkulov$^{2}$, Kur.Myrzakul$^{2}$\thanks{%
Institute of Mathematics, National Academy of Sciences, Alma-Ata,
Kazakhstan}%
, R.Myrzakulov$^{1,2}$ \\
%EndAName
$^{1}$Dipartimento di Fisica, Universita di Lecce and Sezione INFN, 73100
Lecce, Italy\\
$^{2}$Institute of Physics and Technology, 480082 Alma-Ata-82, Kazakhstan}
\date{\today}
\maketitle

\begin{abstract}
We conjecture that many (maybe all) integrable systems and spin systems in
2+1 dimensions can be obtained from the (2+1)-dimensional
Gauss-Mainardi-Codazzi and Gauss-Weingarten equations, respectively. We
also
show that the (2+1)-dimensional Gauss-Mainardi-Codazzi equation which
describes the deformation (motion) of surfaces is the exact reduction of
the
Yang-Mills-Higgs-Bogomolny and Self-Dual Yang-Mills equations. On the
basis
of this observation, we suggest that the (2+1)-dimensional
Gauss-Mainardi-Codazzi equation is a candidate to be integrable and the
associated linear problem (Lax representation) with the spectral parameter
is presented.
\end{abstract}

%\tableofcontents

\section{ Introduction}

$\qquad$Several nonlinear phenomena in physics, modeled by the nonlinear
differential equations (NDE), can describe also the evolution of surfaces
in
time. The interaction between differential geometry of surfaces and NDE
has
been studied since the 19th centure. This relationship is based on the
fact
that most of the local properties of surfaces are expressed in terms of
NDE.
Since the famous sine-Gordon and Liouville equations, the interrelation
between NDE of the classical differential geometry of surfaces and modern
soliton equations has been studied by various points of view in numerous
papers. In particular, the relationship between deformations of surfaces
and
integrable systems in 2+1 dimensions were studied by several authors
[1-14].

The self-dual Yang-Mills equation (SDYME) is a famous example of NDE in
four
dimensions integrable by the inverse scattering method [16-17]. Ward
conjectured that all integrable (1+1)-dimensional NDE may be obtained from
SDYME by reduction [18] (see the book [19] and references therein). More
recently, many soliton equations in 2+1 dimensions have been found as
reductions of the SDYME [20-23].

In this paper we study the deformation of surfaces in the context of its
connection with integrable systems in 2+1 and 3+1 dimensions. We
conjecture
that many integrable (2+1)-dimensional NDE can be obtained from the
deformed
or (2+1)-dimensional Gauss-Mainardi-Codazzi equation (dGMCE) describing
the
deformation (motion) of the surface, as exact particular cases. At the
same
time, integrable isotropic spin systems (SS) in 2+1 dimensions are exact
reductions of the (2+1)-dimensional or in the other words, deformed
Gauss-Weingarten equation (dGWE). This statement is presented as a
conjecture. Also we show that the dGMCE is the exact reduction of two
famous
multidimensional integrable system, namely, the Yang-Mills-Higgs-Bogomolny
equation (YMHBE) and the SDYME.

\section{Fundamental facts on the theory of surfaces}

$\qquad$Let us consider a smooth surface in $R^{3}$ with local coordinates
$x
$ and $t,$ where ${\bf r}(x,t)$ is a position vector. The first and second
fundamental forms of this surface are given by
$$
I=d{\bf r}^{2}=Edx^{2}+2Fdxdt+Gdt^{2}, \qquad II=d{\bf r}\cdot {\bf n}%
=Ldx^{2}+2Mdxdt+Ndt^{2}\eqno(1)
$$
where by defintion $E={\bf r}_{x}^{2},\quad F={\bf r}_{x}\cdot {\bf r}%
_{t},\quad G={\bf r}_{t}^{2}$ and $L={\bf r}_{xx}^{{}}\cdot {\bf n},\quad
M=%
{\bf r}_{tx}\cdot {\bf n},\quad N= {\bf r}_{tt}\cdot {\bf n}$. The unit
normal vector ${\bf n}$ to the surface is given by ${\bf n}=\frac{{\bf r}%
_{x}\wedge {\bf r}_{t}}{|{\bf r}_{x}\wedge {\bf r}_{t}|}$.  There exist
the
third fundamental form
$$
III=d{\bf n}\cdot d{\bf n}=edx^{2}+2f dxdt+g dt^{2}.\eqno(2)
$$
This form, in contrast to $II$, does not depend on the choice of ${\bf n}$
and contains no new information, since it is expressible in terms of $I$
and
$II$ as
$$
III=2H\cdot II-K\cdot I\eqno(3)
$$
where $K,H$ are the gaussian and mean curvatures, respectively. As is well
known in surface theory, the Gauss-Weingarten equation (GWE) for surface
can be written as
$$
{\bf r}_{xx}=\Gamma _{11}^{1}{\bf r}_{x}+\Gamma _{11}^{2}{\bf r}_{t}+L{\bf
n}%
,\quad {\bf r}_{xt}=\Gamma _{12}^{1}{\bf r}_{x}+\Gamma _{12}^{2}{\bf
r}_{t}+M%
{\bf n}, \quad {\bf r}_{tt}=\Gamma _{22}^{1}{\bf r}_{x}+\Gamma
_{22}^{2}{\bf %
r}_{t}+N{\bf n}\eqno(4a)
$$
$$
{\bf n}_{x}=P_{1}^{1}{\bf r}_{x}+P_{1}^{2}{\bf r}_{t},\qquad {\bf n}%
_{t}=P_{2}^{1}{\bf r}_{x}+P_{2}^{2}{\bf r}_{t}\eqno(4b)
$$
where
$$
\Gamma _{11}^{1} =\frac{GE_{x}-2FF_{x}+FE_{t}}{2g},\quad \Gamma
_{11}^{2}=%
\frac{2EF_{x}-EE_{t}-FE_{x}}{2g},\quad \Gamma
_{12}^{1}=\frac{GE_{t}-FG_{x}}{%
2g}\eqno(5a)
$$
$$
\Gamma _{12}^{2} =\frac{EG_{x}-FE_{t}}{2g},\quad \Gamma _{22}^{1}=\frac{%
2GF_{t}-GG_{x}-FG_{t}}{2g},\quad \Gamma
_{22}^{1}=\frac{EG_{t}-2FF_{t}+FG_{x}%
}{2g}. \eqno(5b)
$$
$$
P_{1}^{1} =\frac{MF-LG}{g},\quad P_{1}^{2}=\frac{LF-ME}{g}, \quad
P_{2}^{1}=%
\frac{NF-MG}{g},\quad P_{2}^{2}=\frac{MF-NE}{g}.\eqno(5c)
$$
Here $g=EG-F^{2}$. Now we introduce the orthogonal basis as ${\bf e}_{1}=%
\frac{{\bf r}_{x}}{\sqrt{E}},\quad {\bf e}_{2}={\bf n},\quad {\bf e}_{3}=%
{\bf e}_{1}\wedge {\bf e}_{2}.$ Hence ${\bf r}_{t}=\frac{F}{\sqrt{E}}{\bf
e}%
_{1}-\sqrt{\frac{g}{E}}{\bf e}_{3}. $ Then the GWE (4) takes the form
$$
\left(
\begin{array}{c}
{\bf e}_{1} \\
{\bf e}_{2} \\
{\bf e}_{3}
\end{array}
\right) _{x}=A\left(
\begin{array}{c}
{\bf e}_{1} \\
{\bf e}_{2} \\
{\bf e}_{3}
\end{array}
\right) ,\qquad \left(
\begin{array}{c}
{\bf e}_{1} \\
{\bf e}_{2} \\
{\bf e}_{3}
\end{array}
\right) _{t}=B\left(
\begin{array}{c}
{\bf e}_{1} \\
{\bf e}_{2} \\
{\bf e}_{3}
\end{array}
\right) \eqno(6)
$$
where
$$
A=\left(
\begin{array}{ccc}
0 & k & -\sigma \\
-k & 0 & \tau \\
\sigma & -\tau & 0
\end{array}
\right) ,\qquad C=\left(
\begin{array}{lll}
0 & \omega _{3} & -\omega _{2} \\
-\omega _{3} & 0 & \omega _{1} \\
\omega _{2} & -\omega _{1} & 0
\end{array}
\right) \eqno(7)
$$
and $k_{g}=k=\frac{L}{\sqrt{E}},\qquad \tau_{g}=\tau =-\sqrt{\frac{g}{E}}%
P_{1}^{2},\qquad k_{n}=\sigma =\frac{\sqrt{g}}{E}\Gamma _{11}^{2}$ and $%
\omega _{1}=-\sqrt{\frac{g}{E}}P_{2}^{2},\qquad \omega
_{2}=\frac{\sqrt{g}}{E%
}\Gamma _{12}^{2},\qquad \omega _{3}=\frac{M}{\sqrt{E}}. $ Here $k_{n},
k_{g}, \tau_{g}$ are called the normal curvature, geodesic curvature and
geodesic torsion, respectively. In the case $\sigma =0$ the first equation
of the GWE (6) coincides in fact with the Frenet equation for the curves.
So,
all that we are doing in the next is true for the motion (deformation) of
curves when $\sigma=0$.

The compatibilty condition for the GWE (6) gives the
Gauss-Mainardi-Codazzi equation (GMCE) as
$$
A_{t}-C_{x}+[A,C]=0\eqno(8)
$$
or in elements
$$
k_{t}=\omega _{3x}+\tau \omega _{2}-\sigma \omega _{1},\quad \tau
_{t}=\omega _{1x}+\sigma \omega _{3}-k\omega _{2},\quad \sigma _{t}=\omega
_{2x}+k\omega _{1}-\tau \omega _{3}.\eqno(9)
$$
We can reformulate the linear system (6) in $2\times 2$ matrix form as
$\phi
_{x}=U\phi ,\quad \phi _{t}=V\phi$, where $U=\frac{1}{2i}\left(
\begin{array}{cc}
\tau & k+i\sigma \\
k-i\sigma & -\tau
\end{array}
\right),\quad V=\frac{1}{2i}\left(
\begin{array}{cc}
\omega_{1} & \omega_{3}+i\omega_{2} \\
\omega_{3}-i\omega_{2} & -\omega_{1}
\end{array}
\right).$

\section{Deformation of surfaces}

$\qquad$Now we would like to consider the deformation of the surface with
respect to $y$. We postulate that such deformation or motion of the
surface
is governed by the system [25]
$$
\left(
\begin{array}{c}
e_{1} \\
e_{2} \\
e_{3}
\end{array}
\right) _{x}=A\left(
\begin{array}{c}
e_{1} \\
e_{2} \\
e_{3}
\end{array}
\right) ,\qquad \left(
\begin{array}{c}
e_{1} \\
e_{2} \\
e_{3}
\end{array}
\right) _{y}=B\left(
\begin{array}{c}
e_{1} \\
e_{2} \\
e_{3}
\end{array}
\right) ,\qquad \left(
\begin{array}{c}
e_{1} \\
e_{2} \\
e_{3}
\end{array}
\right) _{t}=C\left(
\begin{array}{c}
e_{1} \\
e_{2} \\
e_{3}
\end{array}
\right) \eqno(10)
$$
where
$$
A=\left(
\begin{array}{lll}
0 & k & -\sigma \\
-k & 0 & \tau \\
\sigma & -\tau & 0
\end{array}
\right) ,\quad B=\left(
\begin{array}{lll}
0 & \gamma_{3} & -\gamma_{2} \\
-\gamma_{3} & 0 & \gamma_{1} \\
\gamma_{2} & -\gamma_{1} & 0
\end{array}
\right) ,\quad C=\left(
\begin{array}{lll}
0 & \omega _{3} & -\omega _{2} \\
-\omega _{3} & 0 & \omega _{1} \\
\omega _{2} & -\omega _{1} & 0
\end{array}
\right)\eqno(11)
$$
and $\gamma_{j}$ are real functions. The system (10) will be called the
deformed or (2+1)-dimensional GWE (for short, dGWE). We remark that first
and third equations of the system (10) are the equations (6) and $A,B$
coincide with formulas (7). The compatibility conditions of the dGWE (10)
gives the deformed or (2+1)-dimensional GMCE (shortly, dGMCE) of the form
[25]
$$
A_{t}-C_{x}+[A,C]=0\eqno(12a)
$$
$$
A_{y}-B_{x}+[A,B]=0\eqno(12b)
$$
$$
B_{t}-C_{y}+[B,C]=0.\eqno(12c)
$$
As we see, equation (12a) is in fact the GMCE (8). This fact explains why
we
call (12) the deformed or (2+1)-dimensional GMCE. The linear problem (Lax
representation) associated with the system (12) can be written as
$$
\Psi_{z } =\lambda^{2}\Psi_{\bar z}+(F^{-}-\lambda^{2} F^{+})\Psi, \qquad
\Psi_{t}=-i\lambda \Psi_{\bar z} + (C+i\lambda F^{+})\Psi \eqno(13)
$$
where $F^{\pm}=A\pm iB$ and $z=\frac{1}{2}(x+iy), \quad \bar
z=\frac{1}{2}%
(x-iy)$. So we can confirm that the dGMCE (12) is a candidate to be
integrable in the sense that for it there exist the Lax representation
with
the spectral parameter (13). Higher hierarchy of the (2+1)-dimensional
GMCE
(12) can be obtained as the compatibility condition of the linear system
[25]
$$
\Psi_{ z} =\lambda^{2}\Psi_{\bar z}+(F^{-}-\lambda^{2} F^{+})\Psi,\qquad
\Psi_{t}=-i\lambda^{n} \Psi_{\bar z} + \sum_{j=0}^{m}\lambda^{j}F_{j}\Psi.
\eqno(14)
$$

\section{Deformation of surfaces induced by (2+1)-dimensional integrable
systems}

$\qquad$In this section we would like to attract an attention on some
aspects of the relation between the deformation of surfaces and integrable
systems in 2+1 dimensions. Now we make some conjectures.

\subsection{Integrable systems in 2+1 dimensions and the deformed GMCE}

$\qquad${\bf Conjecture 1.} {\it Many (maybe all) integrable systems in
2+1 dimensions are particular reductions of equations (12)} [25].

The well known (2+1)-dimensional integrable systems such as the KP and mKP
equations, the Davey-Stewartson (DS) equation and so on, can be obtained
from the dGMCE (12) as some reductions. For instance, the DS-II equation
can
be obtained from the (2+1)-dimensional GMCE (12) as
$$
A=\sqrt{2}i\lambda \sigma _{3}+\frac{1}{\sqrt{2}}\bar{q}\sigma
^{+}+\frac{1}{%
\sqrt{2}}q\sigma ^{-},\qquad B=-\frac{i\lambda }{\sqrt{2}}\sigma
_{3}+\frac{1%
}{\sqrt{2}}\bar{q}\sigma ^{+}+\frac{1}{\sqrt{2}}q\sigma ^{-}\eqno(15a)
$$
$$
C=-\frac{i}{2}(|q|^{2}+\phi _{y}+3\lambda ^{2})\sigma _{3}-3\lambda
\bar{q}%
\sigma ^{+}-3\lambda q\sigma ^{-},\qquad \sigma ^{\pm }=\sigma _{1}\pm
i\sigma _{2}\eqno(15b)
$$
where using isomorfism $so(3)\cong su(2)$, the matrices $A,B,C$ can be
written in $2\times 2$ form. Substituting (15) into the system (12) after
some algebra we get the DS-II equation [24]
$$
iq_{t}+\frac{1}{2}(q_{xx}-q_{yy})-(|q|^{2}+\phi _{y})q=0,\qquad \phi
_{xx}+\phi _{yy}+2(|q|^{2})_{y}=0.\eqno(16)
$$

\subsection{Integrable SS in 2+1 dimensions and the dGWE}

$\qquad$ The conjecture 1 is true also for integrable SS in
2+1 dimensions. But for isotropic subclass of such SS, the following
conjecture takes places .

{\bf Conjecture 2.} {\it Many (and maybe all) integrable isotropic SS in
2+1
dimensions are particular reductions of equations (10)} [25].

As an example, we consider the isotropic Myrzakulov I (M-I) equation
$$
{\bf S}_{t}=({\bf S}\wedge{\bf S}_{y}+u{\bf S})_{x} \eqno(17a)
$$
$$
u_{x}=-{\bf S}\cdot ({\bf S}_{x}\wedge{\bf S}_{y}) \eqno(17b)
$$
which is integrable. In this case we take the identification ${\bf
e}_{1}=%
{\bf S}$, where ${\bf S}$ is the solution of the M-I equation (17) and $%
k^{2}+\sigma^{2}={\bf S}_{x}^{2}$. Then the M-I equation (17) becomes
$$
{\bf e}_{1t}=({\bf e}_{1}\wedge{\bf e}_{1y}+u{\bf e}_{1})_{x}, \quad
u_{x}=-%
{\bf e}_{1}\cdot ({\bf e}_{1x}\wedge{\bf e}_{1y}). \eqno(18)
$$
Now let us assume
\[
\tau=f_{x}, \qquad \gamma_{1}=f_{y}+u, \qquad
\omega_{1}=f_{t}+\partial^{-1}_{x}(\sigma\omega_{3}- k\omega_{2})
\]
$$
\omega_{2}=-\gamma_{3x}-\gamma_{2}\tau+u\sigma, \quad
\omega_{3}=\gamma_{2x}-\gamma_{3}\tau+uk\eqno(19)
$$
where $f(x,y,t,\lambda)$ is a real function. Taking into account formulas
(19) and after eliminating the vectors ${\bf e}_{2}$ and ${\bf e}_{3} $,
the
system (10) takes the form (18). This means that the M-I equation (17) is
the particular exact reduction of the (2+1)-dimensional GWE (10) with the
choice (19). Similarly, we can show that the other isotropic SS in 2+1
dimensions are the exact reductions of the system (10) so that the
conjecture 2 is true at least for existing known integrable isotropic
(2+1)-dimensional SS [25].

\subsection{Integrable SS as exact reductions of the M-0 equation}

$\qquad$Now let us consider the (2+1)-dimensional isotropic Myrzakulov 0
(M-0) equation
(about our notations, see, i.e., [13-15])
$$
{\bf e}_{1t}=\omega _{3}{\bf e}_{2}-\omega _{2}{\bf e}_{3},\qquad \tau
_{y}-\omega _{1x}={\bf e}_{1}\cdot ({\bf e}_{1x}\wedge{\bf
e}_{1y})\eqno(20)
$$
which sometimes we write in terms of ${\bf S}$ as
$$
{\bf S}_{t}=\theta_{1}{\bf S}_{x}+\theta_{2}{\bf S}_{y},\qquad \tau
_{y}-\omega _{1x}={\bf S}\cdot ({\bf S}_{x}\wedge{\bf S}_{y})\eqno(21)
$$
where $\theta_{j}$ are some real functions. The following conjecture takes
places.

{\bf Conjecture 3.} {\it Many (and maybe all) integrable isotropic SS  in
2+1 dimensions are particular reductions of the (2+1)-dimensional
isotropic
M-0 equation (20)-(21)} [25].

For example, the M-I equation (17) is the particular case of (21) as
$$
\theta_{1}=\frac{\omega_{3}\gamma_{2}-\omega_{2}\gamma_{3}}{%
k\gamma_{2}-\sigma\gamma_{3}}, \quad \theta_{2}= \frac{\omega_{2}k-%
\omega_{3}\sigma}{k\gamma_{2}-\sigma\gamma_{3}}.\eqno(22)
$$

\section{The (2+1)-dimensional GCME as exact reduction of the
Yang-Mills-Higgs-Bogomolny equation}

$\qquad$ One of the most interesting and important integrable equations in
2+1 dimensions is the following Yang-Mills-Higgs-Bogomolny equation
(YMHBE)
[19]
$$
\Phi _{y}+[\Phi ,B]+C_{x}-A_{t}+[C,A]]=0\eqno(23a)
$$
$$
\Phi _{t}+[\Phi ,C]+A_{y}-B_{x}+[A,B]=0\eqno(23b)
$$
$$
\Phi _{x}+[\Phi ,A]+B_{t}-C_{y}+[B,C]=0.\eqno(23c)
$$
The important observation is that the dGMCE (12) is the particular case of
the YMHBE (23). In fact, if in the YMHBE we put $\Phi =0$ then the YMHBE
(23) becomes the dGMCE (12). So we can suggest that the dGMCE is a
candidate
to be integrable as the exact reduction of the integrable equation (23).

\section{The (2+1)-dimensional GCME as exact reduction of the Self-Dual
Yang-Mills equation}

$\qquad$Now we study the relationship between the dGMCE (12) and the SDYM
equation. The SDYME reads as
$$
F_{\mu \nu }=^{*}F_{\mu \nu }\eqno(24)
$$
where $^{*}$ is the Hodge star operator and the Yang-Mills field defined
as $%
F_{\mu \nu }=\frac{\partial A_{\nu }}{x_{\mu }}-\frac{\partial A_{\mu }}{%
x_{\nu }}-[A_{\mu },A_{\nu }]$. Let $x_{\alpha }=z+it,\quad x_{\bar\alpha
}=z-it,\quad x_{\beta }=x+iy,\quad x_{\bar\beta}=x-iy$ be the
null-coordinates in Euclidean space for which the metric has the form $%
ds^{2}=dx_{\alpha}dx_{\bar\alpha}+dx_{\beta}dx_{\bar\beta}$. Now the SDYME
takes the form [16-17,19]
$$
F_{\alpha \beta }=0,\quad F_{\bar\alpha \bar\beta }=0,\quad F_{\alpha\bar
\alpha}+F_{\beta \bar\beta }=0\eqno(25)
$$
where $A_{\alpha}=A_{z}+iA_{t}, \quad A_{\bar\alpha}=A_{z}-iA_{t}, \quad
A_{\beta}=A_{x}+iA_{y}, \quad A_{\bar\beta}=A_{x}+iA_{y}.$ The associated
linear system is [19]
$$
(\partial _{\alpha }+\lambda \partial _{\bar\beta })\Psi =(A_{\alpha
}+\lambda A_{\bar\beta })\Psi, \qquad (\partial _{\beta }-\lambda \partial
_{\bar\alpha})\Psi = (A_{\beta }-\lambda A_{\bar\alpha })\Psi \eqno(26)
$$
where $\lambda$ is the spectral parameter and
$$
\frac{\partial}{\partial x_{\alpha}}=\frac{\partial}{\partial z}-i\frac{%
\partial}{\partial t}, \quad \frac{\partial}{\partial
x_{\bar\alpha}}=\frac{%
\partial}{\partial z}+i\frac{\partial}{\partial t}, \quad
\frac{\partial}{%
\partial x_{\beta}}=\frac{\partial}{\partial x}-i\frac{\partial}{\partial
y}%
, \quad \frac{\partial}{\partial x_{\alpha}}=\frac{\partial}{\partial
x}+i%
\frac{\partial}{\partial t}. \eqno(27)
$$
Our second observation: the dGMCE (12) is the particular reduction of the
SDYME (25). In fact, we consider the following reduction of the SDYME
$$
A_{\alpha }= -iC,\quad A_{\bar\alpha }=iC,\quad A_{\beta }=A-iB,\quad
A_{\bar\beta}=A+iB\eqno(28)
$$
and assume that $A,\quad B,\quad C $ are independent of $z$. In this case,
from the SDYME (25) we obtain the (2+1)-dimensional GMCE (12) in Euclidean
coordinates.

\section{Conclusion}

$\qquad$In this paper, we have considered some deformations or in other
terminology, motions of surfaces. We have shown that the corresponding
dGMCE
is integrable in the sense that the associated linear problem (Lax
representation) exists with the spectral parameter. We conjectured that
many
(maybe all) integrable systems in 2+1 dimensions are some reductions of
the
dGMCE. In particular, as example we proved how the DS-II equation can be
obtained from the dGMCE. Although, all known integrable (2+1)-dimensional
isotropic SS can be obtained from the dGMCE, we have conjectured that such
SS can be obtained from  the dGWE as exact reductions. Finally, we proved
that the dGMCE is the particular case of two famous integrable systems
namely the YMHBE and SDYME. It goes in favour of integrability of the
dGMCE.

\section{Acknowledgments}

$\qquad$One of the authors (R.M.) is grateful to M.J.Ablowitz for useful
conservation on the reduction of SDYME. Also he would like to thank
B.G.Konopelchenko, L.Martina and G.Landolfi for helpful discussions and
thanks G.Soliani for careful  reading of the manuscript. This work was
supported in part by MURST of Italy, INFN-Sezione di Lecce, and INTAS,
grant
99-1782. R.M. is grateful to the Department of Physics,  University of
Lecce
for the kind hospitality.

\end{document}